\documentclass{article}
\usepackage{graphicx}
\usepackage{amsmath}
\usepackage{subfigure}
\usepackage{amsfonts}
\usepackage{amssymb}
\date{}

\begin{document}

\title{Efficient and robust constitutive integrators for single-crystal plasticity modeling}
\author{S.N. Kuchnicki$^{1},$ A.M. Cuiti{\~n}o$^{1},$ R.A. Radovitzky$^{2}$\\$^{1}$Department of Mechanical and Aerospace Engineering, \\Rutgers University, Piscataway, NJ\ 08854 USA\\$^{2}$Department of Aeronautics and Astronautics,\\Massachusetts Institute of Technology\\Cambridge, MA 02139 USA}
\maketitle
\begin{abstract}
Small-scale deformation phenomena such as subgrain formation,
development of texture, and grain boundary sliding require
simulations with a high degree of spatial resolution. When we
consider finite-element simulation of metal deformation, this
equates to thousands or hundreds of thousands of finite elements.
Simulations of the dynamic deformations of metal samples require
elastic-plastic constitutive updates of the material behavior to be
performed over a small time step between updates, as dictated by the
Courant condition. Further, numerical integration of
physically-based equations is inherently sensitive to the step in
time taken; they return different predictions as the time step is
reduced, eventually approaching a stationary solution. Depending on
the deformation conditions, this converged time step becomes short
(~$10^{-9} s$ or less). If an implicit constitutive update is
applied to this class of simulation, the benefit of the implicit
update (i.e., the ability to evaluate over a relatively large time
step) is negated, and the integration is prohibitively slow. The
present work recasts an implicit update algorithm into an explicit
form, for which each update step is five to six times faster, and
the compute time required for a plastic update approaches that
needed for a fully-elastic update. For dynamic loading conditions,
the explicit model is found to perform an entire simulation up to 50
times faster than the implicit model. The performance of the
explicit model is enhanced by adding a subcycling algorithm to the
explicit model, by which the maximum time step between constitutive
updates is increased an order of magnitude. These model improvements
do not significantly change the predictions of the model from the
implicit form, and provide overall computation times significantly
faster than the implicit form over finite-element meshes. These
modifications are also applied to polycrystals via Taylor averaging,
where we also see improved model performance.
\end{abstract}

\section{Introduction}
\label{sec:intro}

The explosion in computing power over the last few decades has heightened the
aspirations of numerical analysts. Simulations previously thought difficult
are now commonplace; those once considered impossible or impractical are now
merely time-consuming. Increased computing power has seen an additional focus
on the behavior of polycrystalline metals, of both face-centered cubic and
body-centered cubic crystal structure.

The basis for modern crystal plasticity is modeling of single crystals.
Averaging techniques for polycrystals rely upon a well-formulated
representation of the single crystal. Alternately, direct numerical simulation
of polycrystals \cite{zhao04} models a particular multi-crystalline material
by tracking the interactions between several single crystals. Both of these
methods for capturing the behavior of polycrystals are computationally
intensive; the goal of the modeler at the constitutive level is thus to
provide the fastest model possible while retaining the underlying physics of
the model (and, by extension, the model accuracy).

The fundamental importance of single crystal modeling is reflected in the
abundance of theories on monocrystalline plasticity in the literature. Such
theories begin to appear in the literature in the early 20$^{th}$ century and
are presented by many investigators, including: Taylor \cite{taylor23},
\cite{taylor38}; Schmid \cite{schmid}; Bishop \cite{bishophill}; Hill
\cite{hill66}; Rice \cite{rice71}; Hutchinson \cite{hutch76}; Asaro
\cite{asarorice77}, \cite{asaro83}; Havner \cite{havner73}, \cite{havner92};
Bassani \cite{bassani}; and Kocks \cite{kocks98}. We seek neither to provide a
comprehensive review of these theories, nor do we wish to provide a comparison
among them. Rather, we will use the update procedure of Cuiti\~{n}o and Ortiz
\cite{cuitort93}, first in an implicit form as originally presented. In the
interest of increasing the computational speed of the model, we will present
an explicit form of this model. Then, in view of the relatively short time
steps allowed by the explicit integration, we arrive at a convergence
condition for the explicit integration. Using this condition, we are able to
devise a subcycling scheme that allows the explicit integration to converge
over larger user-specified time steps. We illustrate these model improvements
with numerical examples, which include the computation time required for each.

\section{Constitutive Framework}
\label{sec:constitutive}

As stated earlier, we are applying the model of Cuiti\~{n}o and Ortiz
\cite{cuitort93} for our work here. We briefly review the fundamentals of this
model here in order to make the current work as self-contained as possible.

Our current model follows the lead of numerous previous authors, including but
not limited to Lee \cite{lee67}, \cite{lee69}; Kratochvil \cite{kratoch};
Green and Naghdi \cite{greennag}; Mandel \cite{mandel}; Nemat-Nasser
\cite{nemnass79}; Onat \cite{onat}; Loret \cite{loret}; and Dafilias
\cite{dafilias}. The underlying assumption common to all these theories is
that the overall deformation gradient $\mathbf{F}$ can be decomposed into an
elastic component $\mathbf{F}^{e}$ and a plastic part $\mathbf{F}^{p},$ as:
\begin{equation}
\mathbf{F=F}^{e}\mathbf{F}^{p}\label{def gradient}%
\end{equation}

The existence of such a multiplicative decomposition implies that there is
some stress-free intermediate configuration which contains the deformation due
to plastic slip only; lattice distortion and rotation are presumed to be
contained in $\mathbf{F}^{e}.$ The plastic deformation gradient is assumed to
be volume-conserving. These assumptions ensure that the decomposition
(\ref{def gradient}) is unique. The deformation power per unit undeformed
volume can thus be written
\begin{equation}
\mathbf{P:\dot{F}=\bar{P}:\dot{F}}^{e}+\mathbf{\bar{\Sigma}:\bar{L}}%
^{p}\label{deform power}%
\end{equation}

where $\mathbf{\bar{P}=PF}^{pT}$ is a first Piola-Kirchhoff stress tensor
relative to the intermediate configuration, and $\mathbf{\bar{\Sigma}=F}%
^{eT}\mathbf{PF}^{pT}$ is a stress measure conjugate to the plastic velocity
gradients on the intermediate configuration, given by $\mathbf{\bar{L}}%
^{p}=\mathbf{\dot{F}}^{p}\mathbf{F}^{pT}.$ The work conjugacy relations
(\ref{deform power}) imply forms for the plastic flow rule and elastic
stress-strain relations, i.e.,
\begin{align}
\mathbf{\bar{L}}^{p}  & =\mathbf{\bar{L}}^{p}\left(  \mathbf{\bar{\Sigma}%
,\bar{Q}}\right) \nonumber\\
\mathbf{\bar{P}}  & =\mathbf{\bar{P}}\left(  \mathbf{F}^{e},\mathbf{\bar{Q}%
}\right) \label{flow and stress-strain}%
\end{align}

where $\mathbf{\bar{Q}}$ represents the appropriate internal variables defined
on the intermediate configuration, subject to appropriate evolution equations
(hardening laws). The most general form of the second of (\ref{flow and
stress-strain}) that is material-frame indifferent is
\begin{equation}
\mathbf{\bar{P}=F}^{e}\mathbf{\bar{S}}\left(  \mathbf{\bar{C}}^{e}\right)
\label{general intermed PK1}%
\end{equation}

where $\mathbf{\bar{C}}^{e}=\mathbf{F}^{eT}\mathbf{F}^{e}$ is the right
Cauchy-Green deformation tensor on the intermediate configuration, and
$\mathbf{\bar{S}=\bar{C}}^{e-1}\mathbf{\bar{\Sigma}}$ is a symmetric second
Piola-Kirchhoff stress tensor on the intermediate configuration. For metals,
we can assume a linear relation between $\mathbf{\bar{S}}$ and the elastic
Lagrangian strain on the intermediate configuration, $\mathbf{\bar{E}}%
^{e}=\left(  \mathbf{\bar{C}}^{e}-\mathbf{I}\right)  /2$ without loss of
generality. Higher-order moduli are available, for example, see Teodosiu
\cite{teodos}.

Rice \cite{rice71} has shown that the formulation of $\mathbf{\bar{L}}^{p}$
used here has the structure
\begin{equation}
\mathbf{\bar{L}}^{p}=\sum_{\alpha}\dot{\gamma}^{\alpha}\mathbf{\bar{s}%
}^{\alpha}\otimes\mathbf{\bar{m}}^{\alpha}\label{plastic vel gradient}%
\end{equation}

where $\dot{\gamma}^{\alpha}$ is the shear rate on slip system $\alpha,$ which
has slip direction $\mathbf{\bar{s}}^{\alpha}$ and normal vector
$\mathbf{\bar{m}}^{\alpha}.$ We follow the usual assumption that these slip
rates depend on stress through the corresponding resolved shear stress
$\tau^{\alpha}$ only, meaning
\begin{equation}
\dot{\gamma}^{\alpha}=\dot{\gamma}^{\alpha}\left(  \tau^{\alpha}%
,\mathbf{\bar{Q}}\right) \label{general gamdot}%
\end{equation}

Peirce, \textit{et al} \cite{pierceasaroneed} and several others have proposed
a power law representation for the slip rates,
\begin{equation}
\dot{\gamma}^{\alpha}=\left\{
\begin{array}
[c]{c}%
\dot{\gamma}_{0}\left(  \frac{\tau^{\alpha}}{g^{\alpha}}\right)  ^{\frac{1}%
{m}},\tau^{a}\geq0\\
0,\text{ otherwise}%
\end{array}
\right.
\end{equation}

where $m$ is a strain-rate sensitivity exponent, $\dot{\gamma}_{0}$ is a
reference strain rate, and $g^{\alpha}$ is the current flow stress on slip
system $\alpha$. As noted in the literature \cite{mcginty}, \cite{ling04},
this formulation returns unrealistic slip strain rates for values of
$\tau^{\alpha}/g^{\alpha}$ much different than unity. Hence, we use the form
of this law presented by Cuiti\~{n}o and Ortiz \cite{cuitort93},
\begin{equation}
\dot{\gamma}^{\alpha}=\dot{\gamma}_{0}\left[  \left(  \frac{\tau^{\alpha}%
}{g^{\alpha}}\right)  ^{\frac{1}{m}}-1\right]  ,\tau^{\alpha}\geq g^{\alpha
}\label{improved power law}%
\end{equation}

where it is assumed that $\dot{\gamma}^{\alpha}=0$ if $\tau^{\alpha}%
<g^{\alpha}.$ Note that we have assumed that slip in any system must be
positive. That is, the combination of direction $\mathbf{\bar{s}}^{\alpha}$
and normal vector $\mathbf{\bar{m}}^{\alpha}$ is taken to be a different
system than the combination of direction $-\mathbf{\bar{s}}^{\alpha}$ and
normal vector $\mathbf{\bar{m}}^{\alpha}.$ Hence, slip will only occur for
$\tau^{\alpha}>g^{\alpha}.$ This modification to the power law representation
removes the singularity often seen at $\tau^{\alpha}=g^{\alpha},$ naturally
introducing the result of zero slip velocity when the resolved shear stress
and flow stress are equal. The hardening relation governing the value of
$g^{\alpha}$ is given as
\begin{equation}
\dot{g}^{\alpha}=\sum_{\beta}h^{\alpha\beta}\dot{\gamma}^{\beta}%
\label{hardening evolution}%
\end{equation}

where $h^{\alpha\beta}$ are the hardening moduli. These hardening moduli are
provided from a statistical analysis based on the analysis of Ortiz and Popov
\cite{ortpop}, with the result
\begin{equation}
h^{\alpha\alpha}=h_{c}\left(  t\right)  \frac{2\left[  \tau^{a}\left(
t\right)  \right]  ^{3}}{\tau_{c}^{3}\left(  t\right)  }\left[  \cosh\left(
\frac{\tau_{c}^{2}\left(  t\right)  }{\left[  \tau^{a}\left(  t\right)
\right]  ^{2}}\right)  -1\right] \label{self-hardening}%
\end{equation}

where
\begin{equation}
\tau_{c}\left(  t\right)  =a\mu b\sqrt{\pi n^{\alpha}\left(  t\right)
},\,\,\,\,\,\,\,\,\,\,\,\,h_{c}\left(  t\right)  =\frac{\tau_{c}\left(
t\right)  }{\gamma_{c}^{\alpha}\left(  t\right)  }%
\label{hardening characteristics}%
\end{equation}

are a characteristic shear stress and plastic modulus, $a$ is a coefficient
(on the order of 0.3), $b$ is the Burgers vector, $\mu$ is the shear modulus,
$n\left(  t\right)  $ is the area density of forest dislocations intersecting
the slip plane of system $\alpha,$ and the characteristic strain $\gamma_{c}$
is dependent upon the Burgers vector, the dislocation area density in system
$\alpha$, and the average distance between point obstacles. The off-diagonal
(cross-hardening) terms $h^{\alpha\beta}$ , $\alpha\neq\beta$ are taken to be
zero. The effect of slip in one system on the hardening characteristics of
another is presumed to be described by the forest dislocation density
$n^{\alpha}.$ Francosi and Zaoui \cite{franzao} determined interaction
parameters $a^{\alpha\beta}$ describing the dependence of the forest obstacles
seen in system $\alpha$ on the dislocation density in system $\beta
,\,\rho^{\beta}:$%
\begin{equation}
n^{\alpha}=\sum_{\beta}a^{\alpha\beta}\rho^{\beta}%
\label{forest dislocation dependence}%
\end{equation}

These parameters $a^{\alpha\beta}$ are available for both FCC\ and BCC
crystals \cite{franzao}, \cite{franc83}. For the case of quasi-static
deformation of an FCC\ crystal, these dimensionless parameters are as follows:

\begin{center}%
\begin{table}[tbp] \centering
\begin{tabular}
[c]{|c|c|}\hline
$a_{0}$ & $8\times10^{-4}$\\\hline
$a_{1}/a_{0}$ & $5.7$\\\hline
$a_{2}/a_{0}$ & $10.2$\\\hline
$a_{3}/a_{0}$ & $16.6$\\\hline
\end{tabular}
\caption{FCC Interaction Coefficients\label{fcccoef}}%
\end{table}%
\end{center}

\subsection{Implicit formulation}
\label{sec:implicit}

This constitutive framework was presented in the context of a fully implicit
update scheme by Cuiti\~{n}o and Ortiz. The general update procedure is to
presume the unknowns of the deformations are the slip strain rates
$\dot{\gamma}^{\alpha},$ and to take as input the updated overall deformation
gradient in the undeformed configuration, $\mathbf{F}_{n+1}.$ This is
accomplished by discretizing the viscosity law and solving it for the resolved
shear stress $\tau_{n+1}^{\alpha},$%
\begin{equation}
\tau_{n+1}^{a}=\psi\left(  \frac{\Delta\gamma^{\alpha}}{\Delta
t},g_{n+1}^{\alpha
}\right) \label{discretized tau}%
\end{equation}

where we note that $g_{n+1}^{\alpha}$ is a function of $\Delta\mathbf{\gamma}%
$, the vector of all $\Delta\gamma^{\alpha},$ through the hardening law
(\ref{hardening evolution}). Since we may write
\begin{equation}
\tau^{\alpha}=\left(  \mathbf{\bar{C}}^{e}\mathbf{\bar{s}}^{\alpha}\right)
^{T}\mathbf{\bar{S}\bar{m}}^{\alpha}\label{resolved shear stress}%
\end{equation}

we find that we can express (\ref{discretized tau}) as a function of the form
\begin{equation}
f^{\alpha}\left(  \Delta\mathbf{\gamma}\right)  =\tau_{n+1}^{\alpha}%
-\psi\left(  \frac{\Delta\gamma^{\alpha}}{\Delta
t},g_{n+1}^{\alpha}\right)
=0\label{minimization function}%
\end{equation}

which may be solved using a Newton-Raphson iteration. This is possible because
the Jacobian matrix of (\ref{minimization function}) may be computed
explicitly \cite{cuitort93}, \cite{ling04}. This implicit approach converges
rapidly, within two or three Newton-Raphson iterations.

\subsection{Explicit formulation}
\label{sec:explicit}

While the implicit formulation exhibits good convergence properties,
it (like most implicit integration schemes) is best suited to a
larger increment in time. Explicit finite-element simulations (the
sort to which we wish to apply our model) inherently use small time
integration steps, limited by the time taken for a wave traveling at
the Rayleigh speed to cross an element. Hence, an explicit
formulation for the update is more appropriate for the application
we intend.

The constitutive framework for the explicit form of the model
remains untouched. Additionally, the unknowns of the crystal
plasticity problem remain unchanged. That is, we design our
constitutive update to take $\mathbf{F}_{n+1}$ as an input, solving
for the unknown slip shear rates. Instead of solving iteratively for
the ensemble of slip rates, we determine the shear rates in a
sequential manner. The evaluation at step $t_{n+1}$ is based on the
hardening information and slip velocities from step $t_{n}.$ The
explicit procedure used here can be summarized as follows:

\begin{enumerate}
\item  Calculate $g^{\alpha},$ $h^{\alpha\alpha}$ for all systems based on
step $t_{n}.$

\item  Compute $\mathbf{F}^{e}=\mathbf{F}_{n+1}\mathbf{F}_{n}^{p}$ and
evaluate $\tau^{\alpha}$ for all systems.

\item  If $\tau^{\alpha}<g^{\alpha}$ for all systems, go to step 6. Otherwise:

\item  Apply $\Delta\mathbf{F}^{p}$ $=\dot{\gamma}_{n}^{\alpha}\Delta t\left(
\mathbf{\bar{s}}^{\alpha}\otimes\mathbf{\bar{m}}^{\alpha}\right)  $ due to the
unused slip system $\alpha$ for which $\tau^{\alpha}-g^{\alpha}$ is largest.

\item  Premultiply $\mathbf{F}^{p}$ by $\Delta\mathbf{F}^{p},$ return to step
2 using this result.

\item  Compute new slip rates $\dot{\gamma}_{n+1}^{\alpha}$ and hardening
moduli $h^{\alpha\alpha}$.
\end{enumerate}

Instead of iterating on the ensemble of slip systems, we now
activate the system having the largest overstress and repeat until
no unused systems exist for which the resolved shear stress exceeds
the flow stress from the previous step. We are using the previous
step as a predictor for the state during the next step. Logically,
then, we expect the quality of the prediction to diminish as we
attempt to take longer time steps $\Delta t,$ which is true. As we
will see in Section \ref{sec:accuracy}, the time required for an
integration approaches that used for a fully elastic step. Depending
on the slip activity during step $t_{n},$ the iteration will diverge
for values of $\Delta t$ that are excessively large. Simulations on
finite element meshes have shown that this value of $\Delta t$ may
be smaller than the maximum $\Delta t$ allowed by the mesh. In other
words, the maximal time step achieved by such a simulation would
depend not upon the chosen mesh, but on the material model. We found
this result to be unacceptable, and devised a solution described
below.

\subsection{Subcycling formulation}

In order to work around the maximal time step limitation of the explicit
model, we first need to understand why the model fails. The crystal plasticity
model we have chosen requires that $\tau^{\alpha}\geq g^{\alpha}$ for all
active slip systems $\alpha.$ Further, since the explicit formulation uses the
hardening data from step $t_{n}$ to predict the slip rates at $t_{n+1},$ we
may write
\begin{equation}
\tau_{n+1}^{\alpha}=\tau\left(  \mathbf{\dot{\gamma}}_{n}\mathbf{,}\Delta
t,\mathbf{F}_{n}^{p}\right) \label{explicit tau}%
\end{equation}

That is, the resolved shear stress depends only on the slip increments, the
time step, and the plastic deformation gradient at time $t_{n}.$ This
dependence is manifested through the determination of $\Delta\mathbf{F}^{p}$
described in step 4 of the procedure above. The change in the flow stress on a
system $\alpha$ is evaluated as
\begin{equation}
g_{n+1}^{\alpha}=g_{n}^{\alpha}+h_{n}^{\alpha\alpha}\dot{\gamma}_{n}^{\alpha
}\Delta t\label{gn+1}%
\end{equation}

which is linear in $\Delta t.$ The relation (\ref{explicit tau}) depends only
on $\Delta t,$ albeit nonlinearly. Since we require $\tau^{\alpha}\geq
g^{\alpha},$ we can set (\ref{explicit tau}) equal to (\ref{gn+1}) and solve
iteratively for $\Delta t_{c},$ where we define $\Delta t_{c}$ as the maximum
(``critical'') time increment for which our constraint is satisfied.

In practice, we do not need the actual value of $\Delta t_{c}$ in
order to proceed with the integration. We are merely interested in
whether we need to invoke the subcycling algorithm. If the given
$\Delta t<\Delta t_{c},$ the evaluation may proceed without
subcycling. If $\Delta t>\Delta t_{c},$ subcycling is activated. It
should be noted that the increments used for the deformation
gradient within the subcycling procedure must be consistent with the
multiplicative decomposition. The integration
proceeds by dividing the overall step $\Delta t$ into smaller
increments $\Delta \tilde{t} = \frac{\Delta t}{n} \leq \Delta
t_{c}$, where $n$ is the number of subcycles. To summarize, the
procedure we use is this:

\begin{enumerate}
\item  Test the condition $\tau^{\alpha}\geq g^{\alpha}$ for all systems
active at time $t_{n+1}.$ If true, go to step 4.

\item  Evaluate the desired $\Delta\mathbf{F.}$ Bisect the desired time step,
test this step applying $\Delta\mathbf{F}^{inc}=\left(  \Delta\mathbf{F}%
\right)  ^{1/2}.$

\item  Repeat until condition in step 1 is satisfied.

\item  Exit, returning the number of subcycles $n,$ the time step
$\Delta\tilde{t}=\Delta t/n$, and the incremental deformation $\left(
\Delta\mathbf{F}\right)  ^{1/n}.$
\end{enumerate}

This procedure has the effect of adding more evaluations of the
explicit step for each global step, since the convergence test
itself is an explicit evaluation. However, we will see from the
relative computational speed of the models that this small net loss
of time is converted into a large gain. In general, not every step
of a simulation requires the largest amount of subcycles to
complete. Without subcycling, we would be constrained by the
\textit{smallest }$\Delta t_{c}$ required by any step. The
subcycling implementation takes advantage of the larger $\Delta
t_{c}$ values available in most steps. Additionally, for a
large-scale simulation, not every quadrature point in the mesh
requires subcycling at a given time step. It may well be, for a
dynamic simulation, that part of the material is highly plastic
while another is still in the elastic range. The subcycling
implementation allows the steps that are easier to evaluate to be
finished quickly, without wasting many steps moving through the less
computationally-intensive elastic region. In this way, the
subcycling algorithm is analogous to an adaptive time step that
activates only when necessary.

Note that, in actuality, we never solve for the critical time step $\Delta
t_{c}.$ Instead, we solve for the largest step $\Delta t/2^{n}$ such that the
condition in step 1 is satisfied. Our chosen method has two useful benefits.
First, bisection is nearly trivial to implement and is computationally
inexpensive (no derivatives to evaluate, for example). Second, while we do not
arrive at the critical time step, we return a time step $\Delta\tilde{t}%
\leq\Delta t_{c}.$ Since $\Delta t_{c}$ is the maximum step we can
take while maintaining convergence, the method we have chosen
provides a built-in error tolerance to the time step evaluation. If
we try to solve for $\Delta t_{c}$ exactly, we may find that small,
unavoidable numerical errors in this solution would cause the
integration to fail. The bisection solution helps reduce or
eliminate this pitfall, resulting in a more robust crystal
plasticity algorithm.

\section{Accuracy of the Algorithm}
\label{sec:accuracy}

We will evaluate our explicit model and subcycling implementation in two
phases. First, we will compare the results from the implicit model to the
explicit, with the goal of proving that the computational speed we gain does
not come at too steep of a price in accuracy. Then, we will apply the
algorithm to some larger-scale simulations, highlighting the increase in
processing speed seen in using the subcycling algorithm.

\begin{figure}
[ptb]
\begin{center}
\includegraphics[width=.60\textwidth,height=!]{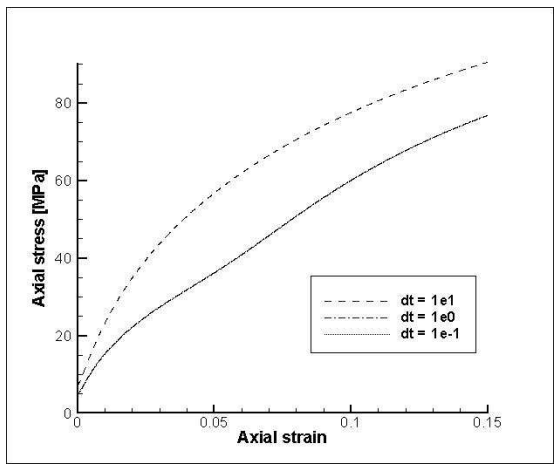}%
\caption{Variations in implicit model results with time step.}%
\label{implicit timing}%
\end{center}
\end{figure}

An important consideration in comparing the results from the implicit and
explicit algorithms is time-convergence (stationarity) of the results,
especially for the implicit model. Most numerical integrations exhibit some
form of time-step dependency; Figure \ref{implicit timing} illustrates the
dependency for several simulations of the quasi-static deformation of a copper
crystal in tension along the [112] crystal axis. If we choose a time step
$\Delta t=10,$ the model predicts a different response than $\Delta t=1$ or
$\Delta t=0.1.$ Note that the predicted responses for the latter two time
steps differ by 0.008\% of the smaller value. If we require our model
responses to be stationary for smaller time steps, then we should use $\Delta
t$ of no larger than 1. Since we wish to apply our models to dynamic
finite-element simulations using very small time steps, i.e., $O\left(
10^{-8}\right)  ,$ stationarity of the results with respect to the time step
is a necessary condition. The results presented here match well with
experimental data \cite{franzao}; we list the model parameters used in Table
\ref{Table 1}.%

\begin{center}%

\begin{table}[tbp] \centering

\begin{tabular}
[c]{|c|c|}\hline
Elastic Constant $C_{11}$ & $168.4$ GPa\\\hline
Elastic Constant $C_{12}$ & $121.4$ GPa\\\hline
Elastic Constant $C_{44}$ & $75.4$ GPa\\\hline
$g_{0}$ & $2.0$ MPa\\\hline
$\dot{\gamma}_{0}$ & $10$ $\sec^{-1}$\\\hline
$m$ & $0.1$\\\hline
$a$ [see Eq. (\ref{hardening characteristics})] & $0.3$\\\hline
$b$ & $2.56\times10^{-10}$\\\hline
Initial dislocation density, $\rho_{0}$ & $10^{12}$ $m^{-2}$\\\hline
Saturation dislocation density, $\rho_{sat}$ & $10^{15}\,m^{-2}$\\\hline
Saturation strain, $\gamma_{sat}$ & $0.5\%$\\\hline
\end{tabular}
\caption{Definition of Symbols\label{Table 1}}%
\end{table}%
\end{center}

\begin{figure}
[ptb]
\begin{center}
\includegraphics[width=.60\textwidth,height=!]{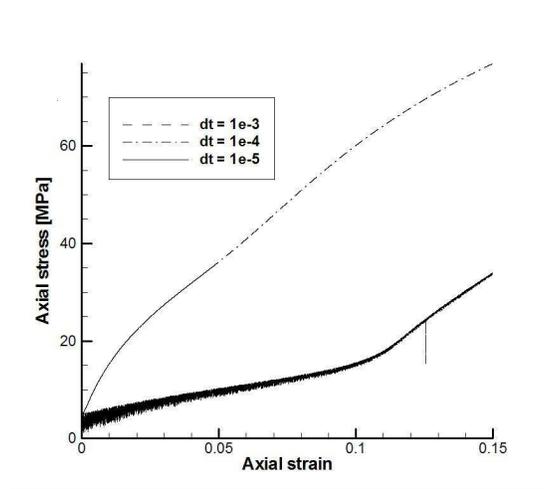}%
\caption{Time-convergence behavior of the explicit model.}%
\label{explicit timing}%
\end{center}
\end{figure}

Figure \ref{explicit timing} is of the same form as Figure \ref{implicit
timing}, but for the explicit model. The time-converged step is around $\Delta
t=10^{-4},$ while the results at $\Delta t=10^{-3}$ are shown to be vastly different.%

\begin{figure}
[ptb]
\begin{center}
\includegraphics[width=.60\textwidth,height=!]{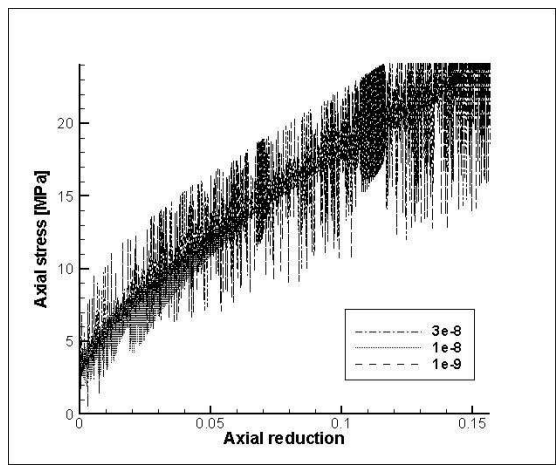}%
\caption{Showing the larger-scale oscillations introduced in the subcycling
response for increased time steps. Larger steps than $\Delta t=3\times10^{-8}$
diverge.}%
\label{subcycling test 2}%
\end{center}
\end{figure}

Our next step is to show that the explicit model with subcycling
gives results similar to the explicit model without subcycling. In
order to illustrate this part, we applied the implicit model and
explicit model with and without subcycling to a
single-integration-point simulation of a rolling test. The crystal
is compressed at high strain rate ($\sim5000/s$) along its [001]
axis, with the global $X-Y$ axes at 45$^{\circ}$ to the crystal
$x-y$ axes. The global $Y$ face is constrained. We choose this sort
of test for two reasons: First, this test is closer to the type of
simulation we wish to perform with the explicit model. Second, we do
not wish for the constitutive tangents to cloud the results, since
we intend to apply our subcycling algorithm to simulations not
requiring the constitutive tangents. Figures \ref{implicit rolling
test}-\ref{subcycling test 2} show the time-convergence behavior for
the implicit, explicit and subcycling models. Figure \ref{full
compare rolling} compares the results for a time-converged implicit
and explicit model to results that include the use of subcycling
($\Delta t\symbol{126}10^{-8}).$ The difference among the models at
15\% reduction is 0.08\% of the smallest value (in this case, the
implicit). If we take a smaller time step with the subcycling model,
the $\Delta t_{c}$ constraint is not violated and the subcycling
model becomes the explicit model exactly. While we see oscillations
in the force-deformation response when we apply subcycling, these
oscillations are in general about the converged solution. As
expected, the subcycling formulation continually over- and
under-predicts the stress response, eventually settling on the
correct value. We gain about an order of magnitude in the time step
using subcycling. If we attempt to take a larger step than what is
depicted, we find that the simulation no longer converges at a very
early point in the deformation (within the first 50 steps).

A closer analysis of the output shows that the subcycling tests that
failed had large oscillations in the stress-strain curve, as
demonstrated by the simulation using $\Delta t=10^{-8}$. This manner
of failure is due to the issue of time-convergence. If the original
input time step $\Delta t$ is greater than the converged time step,
then steps that involve subcycling will be on a time-converged
solution curve, while any steps that do not involve subcycling may
not be on the time-converged curve. The overall solution, then, runs
on two paths; this condition eventually causes the overall solution
to diverge.

While these oscillations are a concern at the integration-point
level, this effect tends to diminish within the framework of a
large-scale finite-element solution. For such simulations, the
subcycling implementation acts to make the system more robust. This
is because of the necessarily smaller time steps used in the
finite-element solution. These smaller time steps are more likely to
be converged\footnote{Certainly, the smaller time step is closer to
the converged step than the single-integration-point steps above}.
The difference between the subcycling path and the smoother path
without subcycling will be less dramatic. Another important point
from this analysis is that the subcycling algorithm used here does
not increase the effective time step over which the model can
progress without bound. There are two reasons for this. The first is
numerical precision; that is, the $\Delta \mathbf{F}^{\frac{1}{n}}$
calculations lead to matrices that approach the identity as
$n\rightarrow\infty.$ For steps of smaller than about $10^{-15},$
the difference between $1$ and $(1+\varepsilon)$ is no longer
computationally resolved. Hence, if we try to take steps smaller
than this, the $\Delta \mathbf{F}$ matrix will be seen by the
machine as the identity. Secondly, our subcycling algorithm uses an
estimate over the entire desired time step before taking the first
step. This estimate will be good so long as the velocity in every
slip system $\left(  \dot{\gamma}^{\alpha}\right) $ decreases across
the number of cycles. If we activate new systems in a subcycle, or
if the velocity in a previously activated system exceeds the initial
estimate, we are no longer guaranteed that the model will converge.
It may be argued, then, that the current subcycling algorithm should
be modified to be recursive, testing both the initial time step and
that of each subcycle. The present formulation, however, has proven
to be sufficient; the benefit gained by introducing a recursive
subcycling algorithm may well be outweighed by the effort required
in its implementation, both in terms of programming and computation
time.

\begin{figure}
[ptb]
\begin{center}
\includegraphics[width=.60\textwidth,height=!]{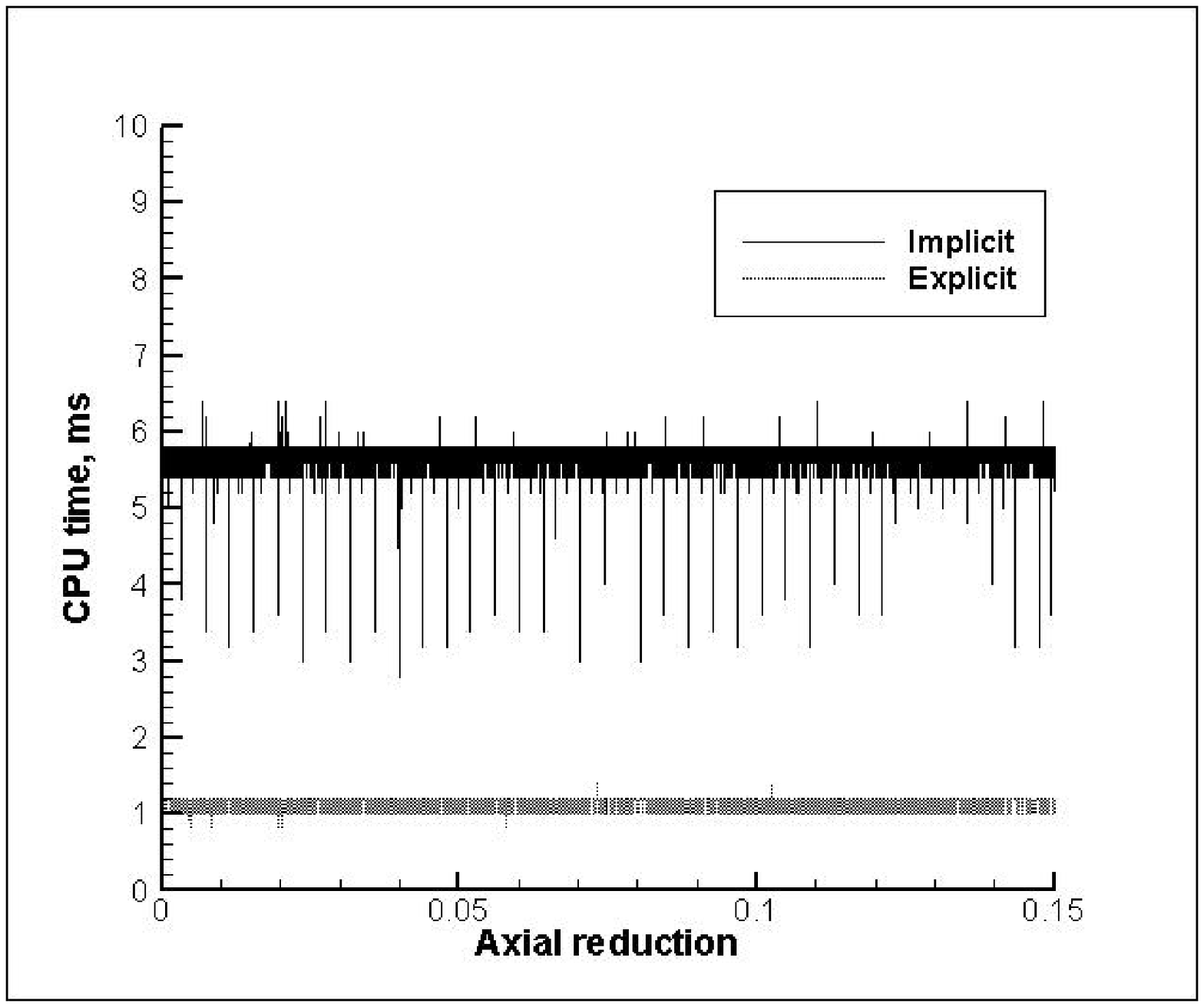}%
\caption{Comparison between compute times per second for the implicit and
explicit update formulations.}%
\label{fcc timing}%
\end{center}
\end{figure}

We close our discussion of the constitutive-level results with comparisons of
the computational time between the implicit and explicit integration. Figure
\ref{fcc timing} presents the variation of computation time for one
integration point versus deformation for the rolling-type test used above. (We
omit the times for the subcycling model since the core of the integration is
the explicit model.) We obtained these curves by running each update over the
same set of parameters 50 times and averaging the processor time used for the
full set. These results show that the explicit calculation is about five to
six times faster than the implicit. While by no means dramatic, this gain of
time is significant; an implicit simulation lasting a week would (at the same
strain step)\ take about a day for the explicit model to complete. Even for
the constitutive-level analysis in this section, we can see the interplay
between the largest stationary time step and the compute speed. The tensile
tests (Figures \ref{implicit timing} and \ref{explicit timing}) at low strain
rate allow the implicit model to take a step four orders of magnitude longer
than the explicit, more than counteracting the half order of magnitude of
compute speed gained by the explicit model. However, the rolling-type tests
presented above require the implicit model to take a time step on the order of
$\Delta t=10^{-10}$ and the explicit model\footnote{Without subcycling.} to
take a time step on the order of $\Delta t=10^{-9}.$ In this case, the
explicit model is simply faster, by an overall factor of about 50.%

\begin{figure}
[ptb]
\begin{center}
\includegraphics[width=.60\textwidth,height=!]{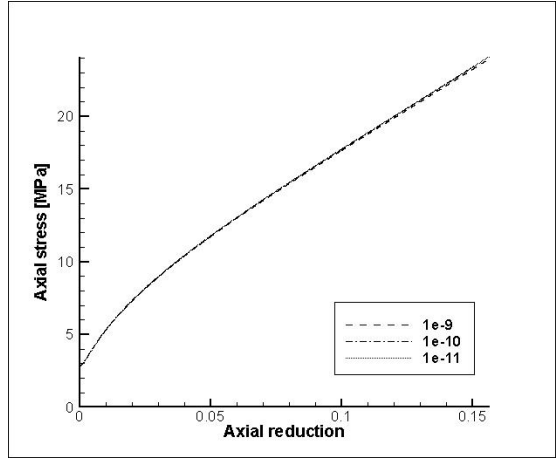}%
\caption{High-rate test using the implicit formulation. The results using
$\Delta t=10^{-10}$ and $\Delta t=10^{-11}$ differ by 0.06\% at 15\%
reduction.}%
\label{implicit rolling test}%
\end{center}
\end{figure}

\begin{figure}
[ptb]
\begin{center}
\includegraphics[width=.60\textwidth,height=!]{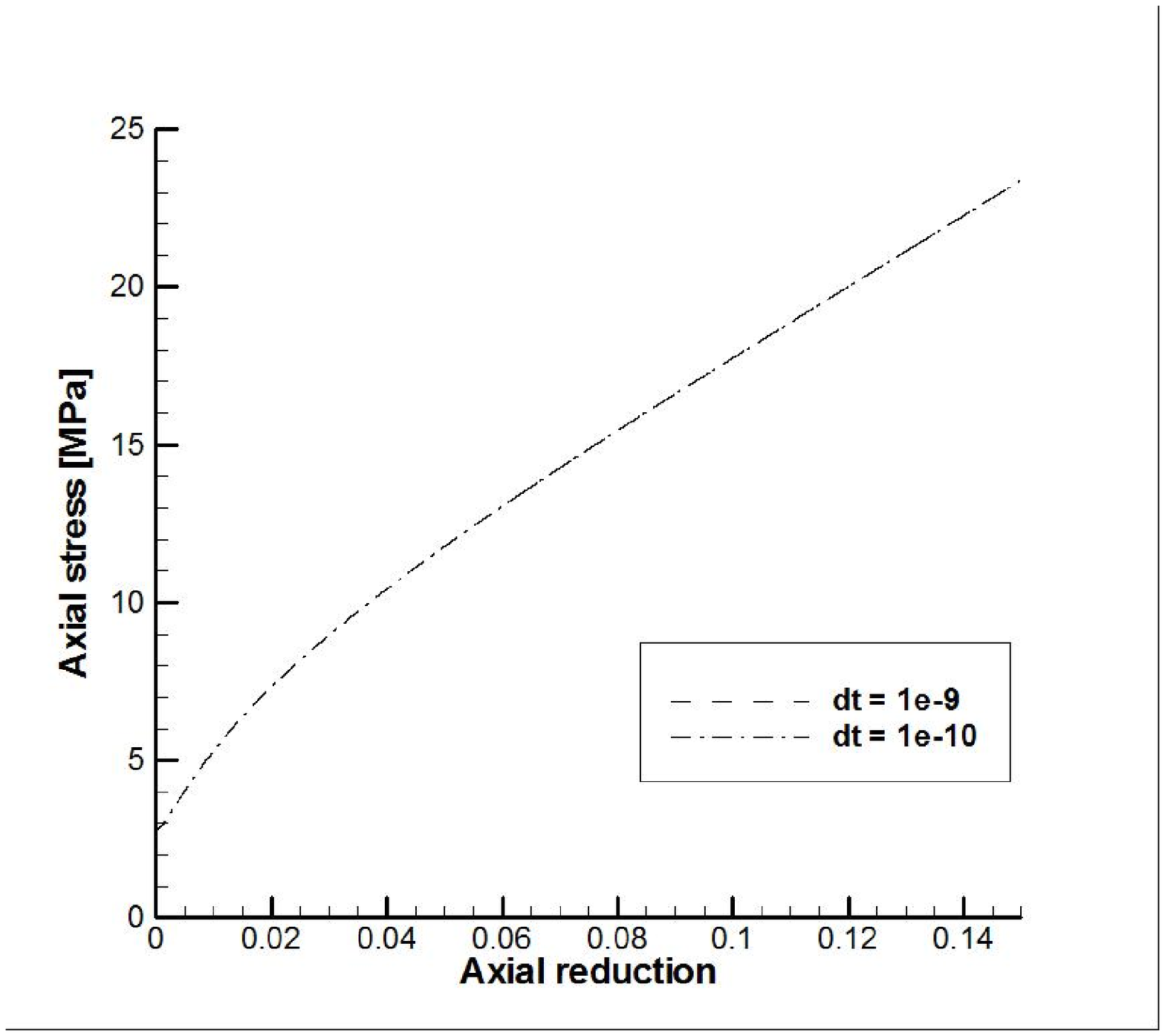}%
\caption{Time stationarity test for compressive deformation using the explicit
model. The two curves here differ by 0.0009\% at 15\% reduction.}%
\label{explicit rolling test}%
\end{center}
\end{figure}

\begin{figure}
[ptb]
\begin{center}
\includegraphics[width=.60\textwidth,height=!]{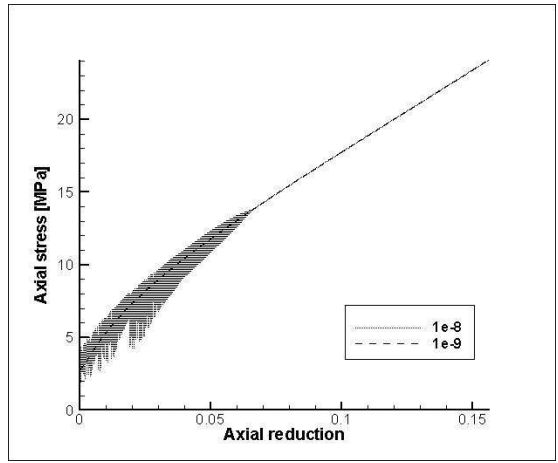}%
\caption{Convergence test using subcycling. After the initial oscillations,
the curves shown here are exactly atop one another.}%
\label{subcycle rolling 1}%
\end{center}
\end{figure}

\begin{figure}
[ptb]
\begin{center}
\includegraphics[width=.60\textwidth,height=!]{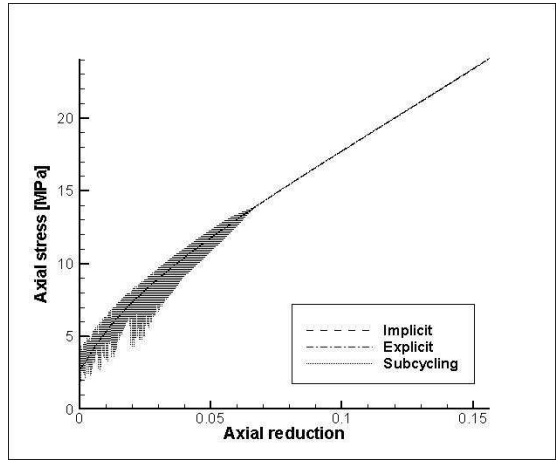}%
\caption{Converged curves for implicit, explicit and subcycling models. }%
\label{full compare rolling}%
\end{center}
\end{figure}

\section{Large-Scale Simulation Results}

The motivation for our foregoing discussion was the application of the derived
material models to large-scale simulations requiring many integrations over
small time steps. Here we provide a few examples of such tests performed using
these algorithms.

\subsection{Rolling Test of FCC\ Aluminum}

We begin with a 1 mm aluminum cube oriented such that the crystal
$z$ axis lies on the global $Z$ axis, and the other two axes defined
by a 45$^{\circ} $ rotation about the crystal $z$ axis. This cube is
subjected to a strain rate of $\sim5000/s$ along the global $Z$
axis. The cube is constrained so that expansion in the global
$Y\;$direction is not possible, but it may deform freely in all
other directions. The results below show the force applied versus
the net reduction, and the initial and final textures. The
qualitative behavior of the force-deformation curve agrees with the
physical interpretation of the simulation. The simulation captures
the fluctuations in the applied force due to wave reflection from
the opposite face of the cube, which are eventually damped out. To
help make the case for our subcycling algorithm, we present results
for the explicit model both with and without our subcycling
algorithm.

\begin{figure}
[ptb]
\begin{center}
\includegraphics[width=.60\textwidth,height=!]{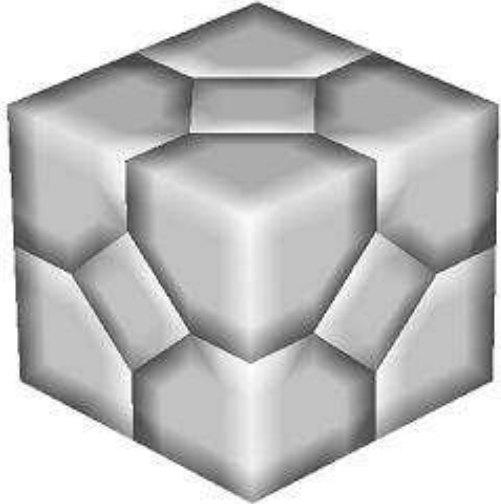}%
\caption{Initial cube for the tests that follow. This cube is meshed with 384
tetrahedra. The heavy lines denote boundaries between twenty-sided polyhedral grains.}%
\label{initmesh}%
\end{center}
\end{figure}

The effects of adding subcycling to the explicit model are striking. We will
use an integration over a tetrahedral mesh of a cubic sample, modeling one
grain in each direction (initial shape given in Figure \ref{initmesh})
deformed in a rolling experiment as a sample case. Without the addition of
subcycling, a simulation using the explicit integration model described above
for an FCC material would require eight hours before the thickness reaches
50\% reduction, using eight processors. This is not due to a limitation of the
mesh, but rather of the material model; the material model is found to require
a stable time step of about 25\% that required by the tetrahedron geometry. If
we try the identical simulation using the subcycling algorithm described
above, we find that the maximum stable time step becomes that of the
mesh - the material model is no longer the limiting factor. As a result,
this same simulation on the same eight processors requires about fifteen
minutes to reach 50\% reduction. The gain of time is nonlinear because not all
quadrature points require subcycling at any one time. It may be that only one
of the approximately 380 integration points requires a smaller step at a given
time. Without a subcycling algorithm, that one integration point would be
enough to require the time step for the entire mesh to be reduced by a factor
of four or more.%

\begin{figure}
[ptb]
\begin{center}
\includegraphics[width=.60\textwidth,height=!]{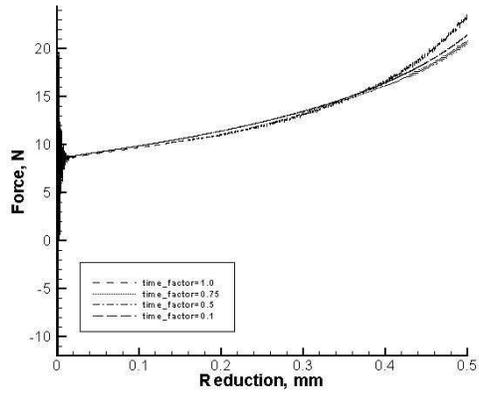}%
\caption{Comparison of force-deformation curves for explicit model runs with
several different time steps. }%
\label{1x1x1force}
\end{center}
\end{figure}

Figure\textit{\ }\ref{1x1x1force}\textit{\ }compares the force-reduction
curves for three simulations. The difference among these simulations is the
maximum allowed time step. The time\_factor variable in the plot legend is a
premultiplying factor applied to the mesh stable time step. The simulation
using time\_factor 0.1 required no subcycling to complete. At first glance, it
seems that the simulation without subcycling is capturing more of the
reflected waves than the one using subcycling at the beginning of the
deformation. However, if we look at a close-up of this region, we find that
the simulations cross the same points where they exist; Figure \ref{1x1x1force
close-up}. That is, we are capturing the same curve with varying levels of
detail. This implies that the additional oscillations seen for
time\_factor=0.1 are a result of having a smaller time step, and not a failing
of the subcycling implementation.

\begin{figure}
[ptb]
\begin{center}
\includegraphics[width=.60\textwidth,height=!]{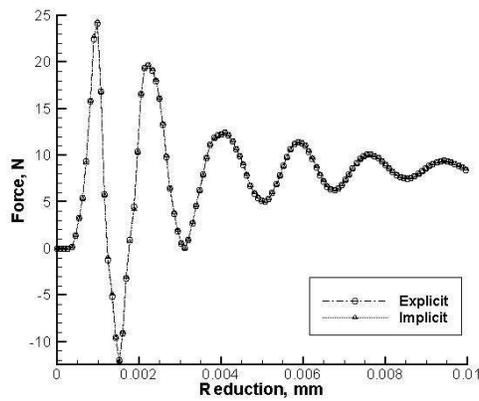}%
\caption{Close-up of the implicit and explicit force curves. Note that the
symbols lie atop one another.}%
\label{implicit explicit cu}%
\end{center}
\end{figure}

\begin{figure}
[ptb] \subfigure[Implicit]{
\includegraphics[width=.45\textwidth, height=!] {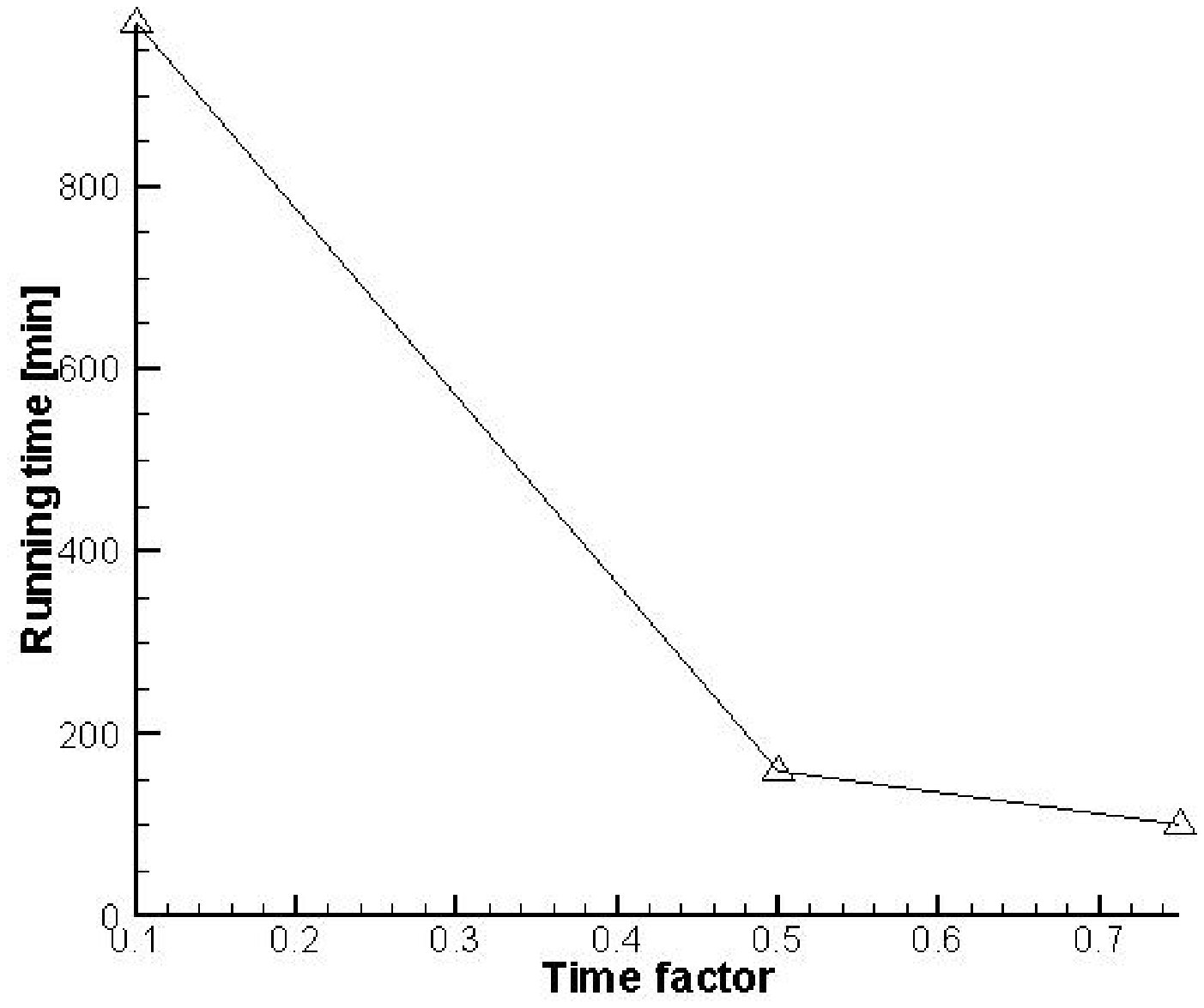}%
} \subfigure[Explicit] {
\includegraphics[width=.45\textwidth, height=!] {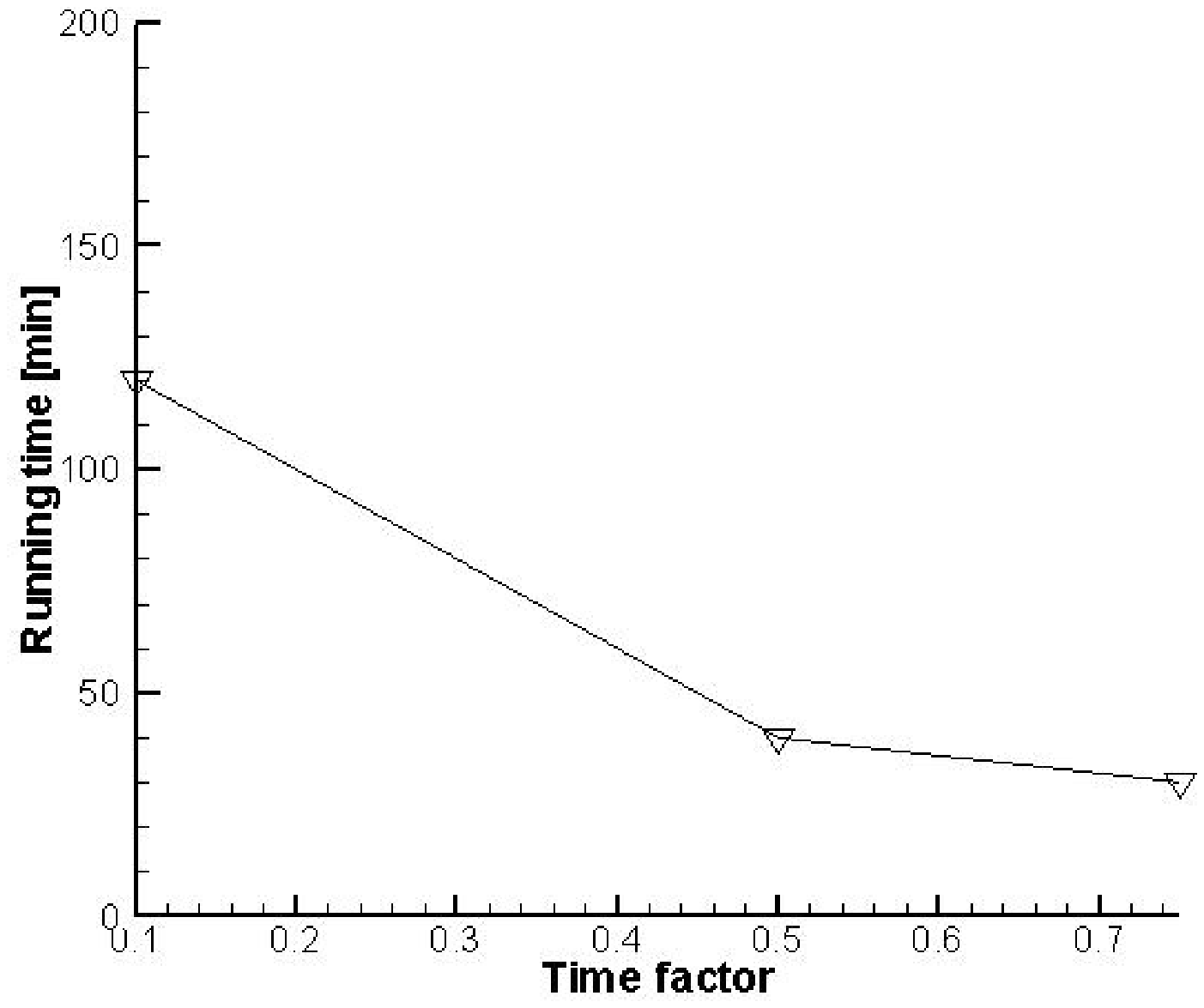} 
}
\caption{Comparison of running time versus time step for both implicit and explicit models.}%
\label{running times}%
\end{figure}

We also attempted a simulation of this deformation using our implicit
formulation. The force-deformation results are shown in Figure \ref{implicit
explicit} and \ref{implicit explicit cu}, compared to the explicit formulation
at time\_factor of 0.5. (Note that the implicit model was capable of stepping
as large as time\_factor 0.75, but the results at this time step are not
stationary as defined above). We plot the time taken by each model versus the
time step in Figure \ref{running times}. Interestingly, we start to see a
negative return when we increase the time step past 75\% of the mesh time.
This is due to the larger number of subcycles required for the full mesh time
step to proceed. The number of extra evaluations required outweighs the 33\%
larger time step for this case. In fact, the simulation with time\_factor=1.0
required almost as much time to complete as that with time\_factor=0.5. Thus,
blindly increasing the time step up to the value allowed by the mesh does not
necessarily produce the most efficient simulation. Note also that the implicit
model is slower at every time step by about a factor of 5; this coincides
nicely with the predictions of Figure \ref{fcc timing}.

\begin{figure}
[ptb] \centering \subfigure[$1.0$.]{
\includegraphics[width=.45\textwidth,height=!] {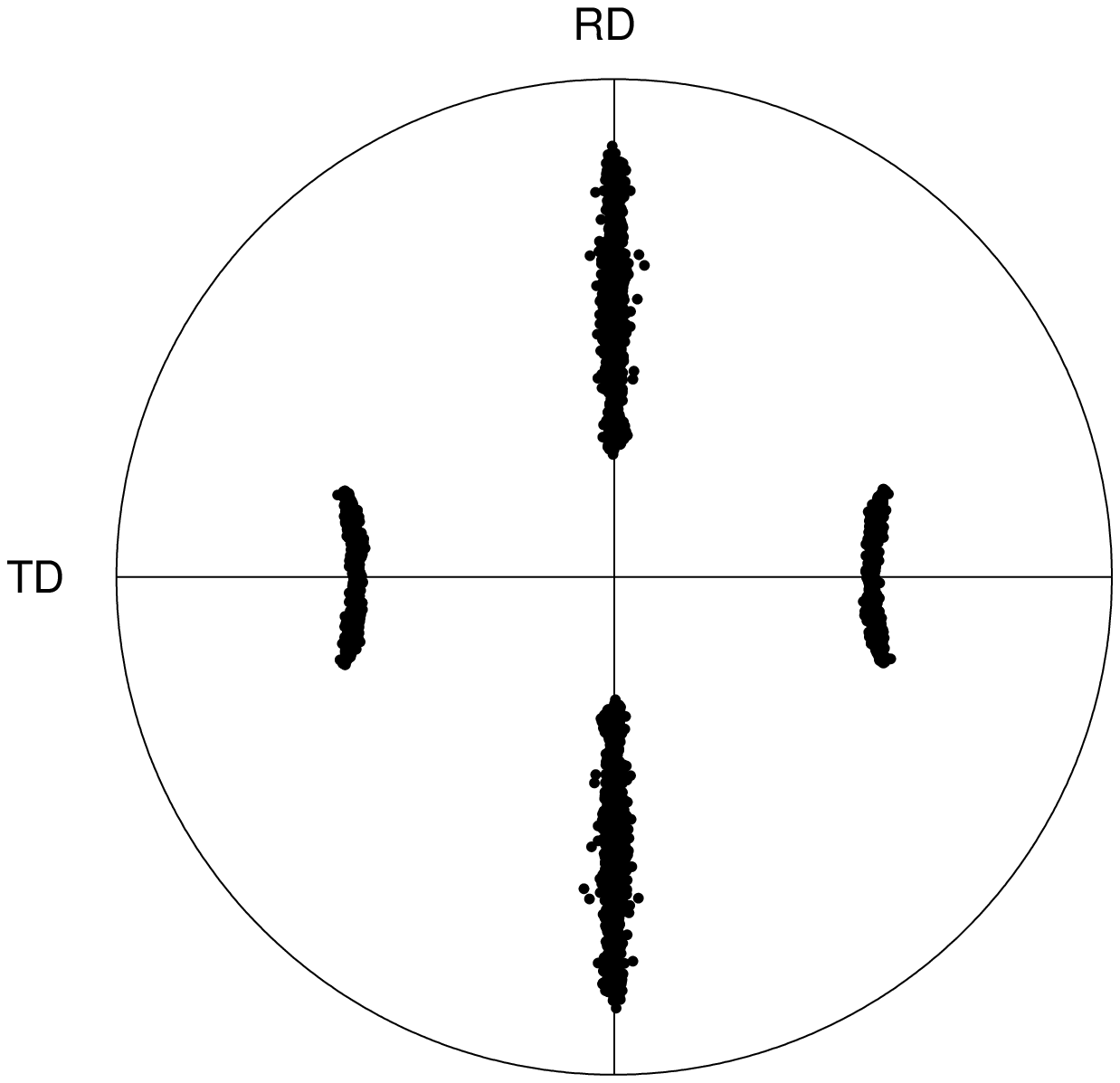}
} 
\subfigure[$0.75$.]{
\includegraphics[width=.45\textwidth,height=!] {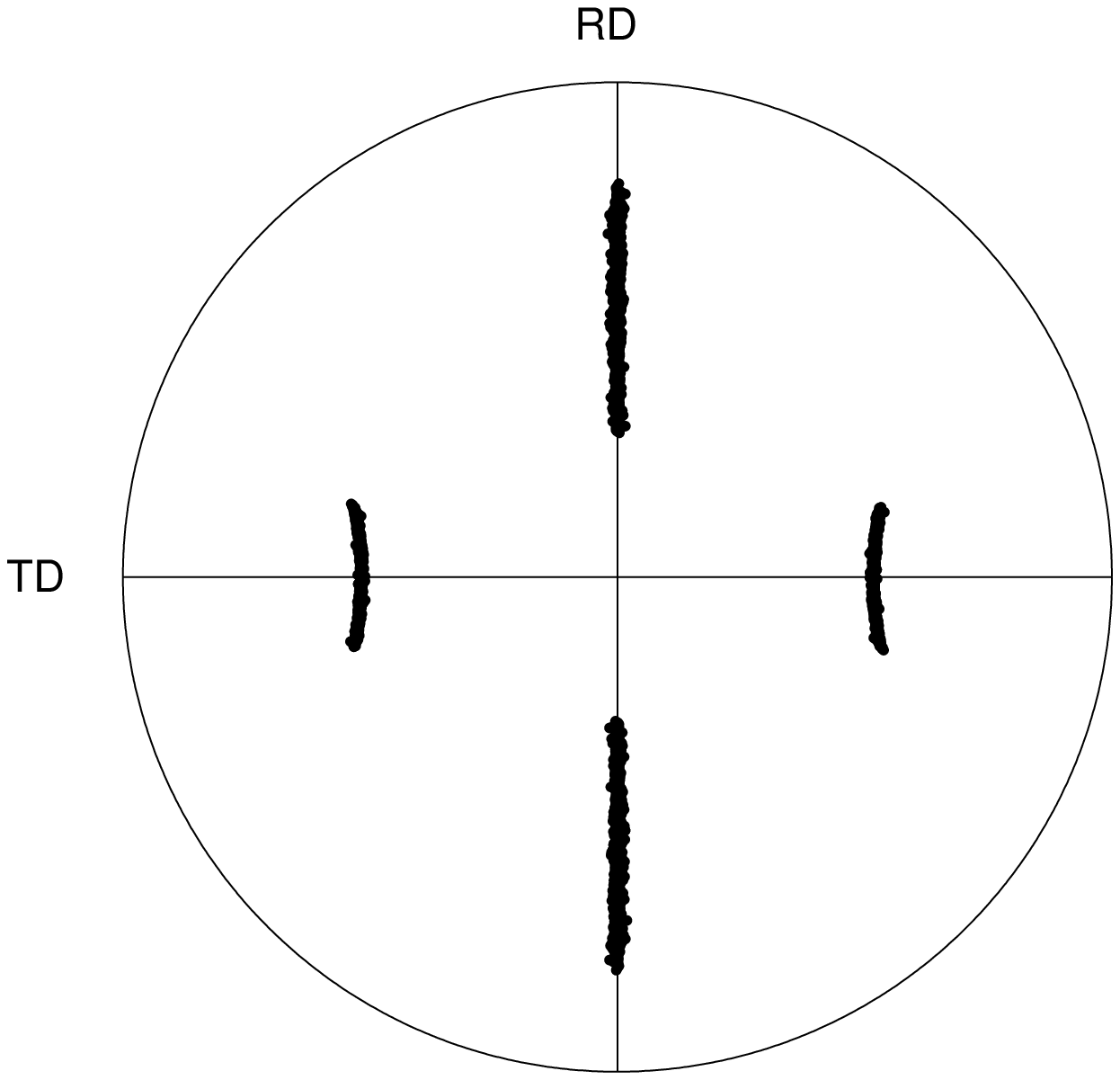}
} 
\vspace{.7in} 
\subfigure[$0.50$.]{
\includegraphics[width=.45\textwidth,height=!]{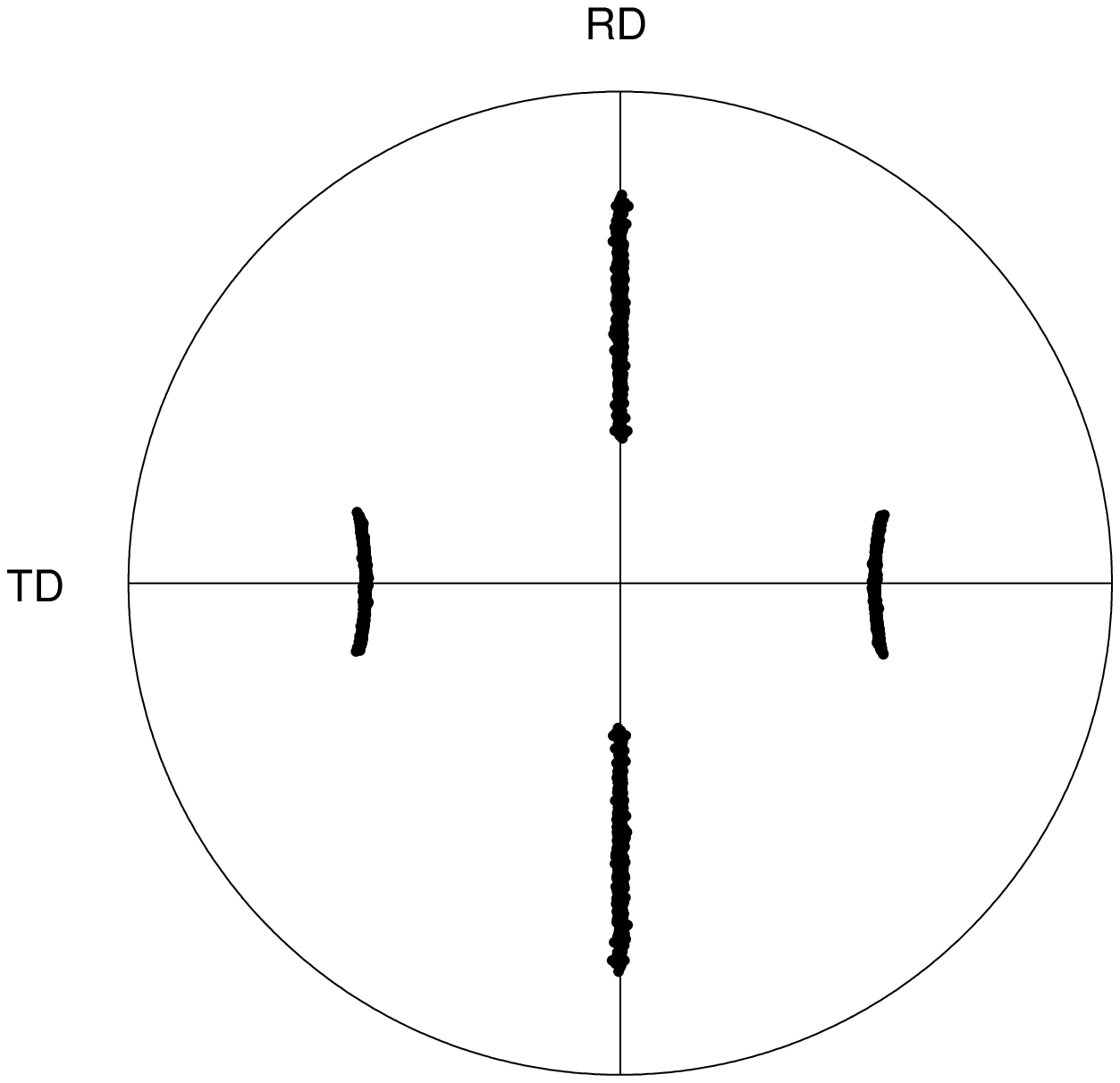}
} 
\subfigure[$0.10$.] {
\includegraphics[width=.45\textwidth,height=!] {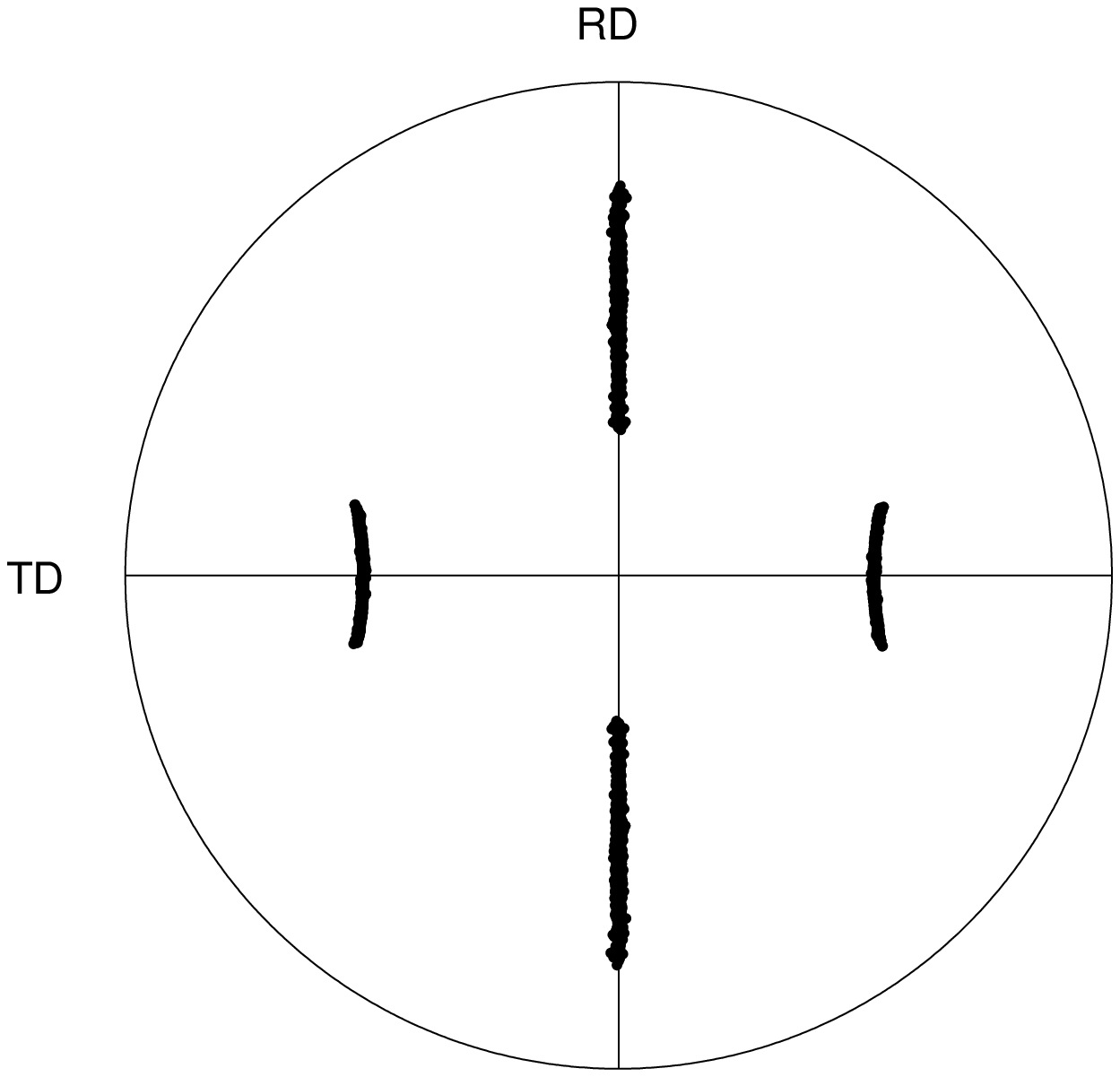}
}
\caption{Final textures for the explicit model simulations for
different time factors.} \label{final explicit textures}
\end{figure}

\begin{figure}
[ptb]
\subfigure[0.50]
{
\includegraphics[width=.45\textwidth, height=!] {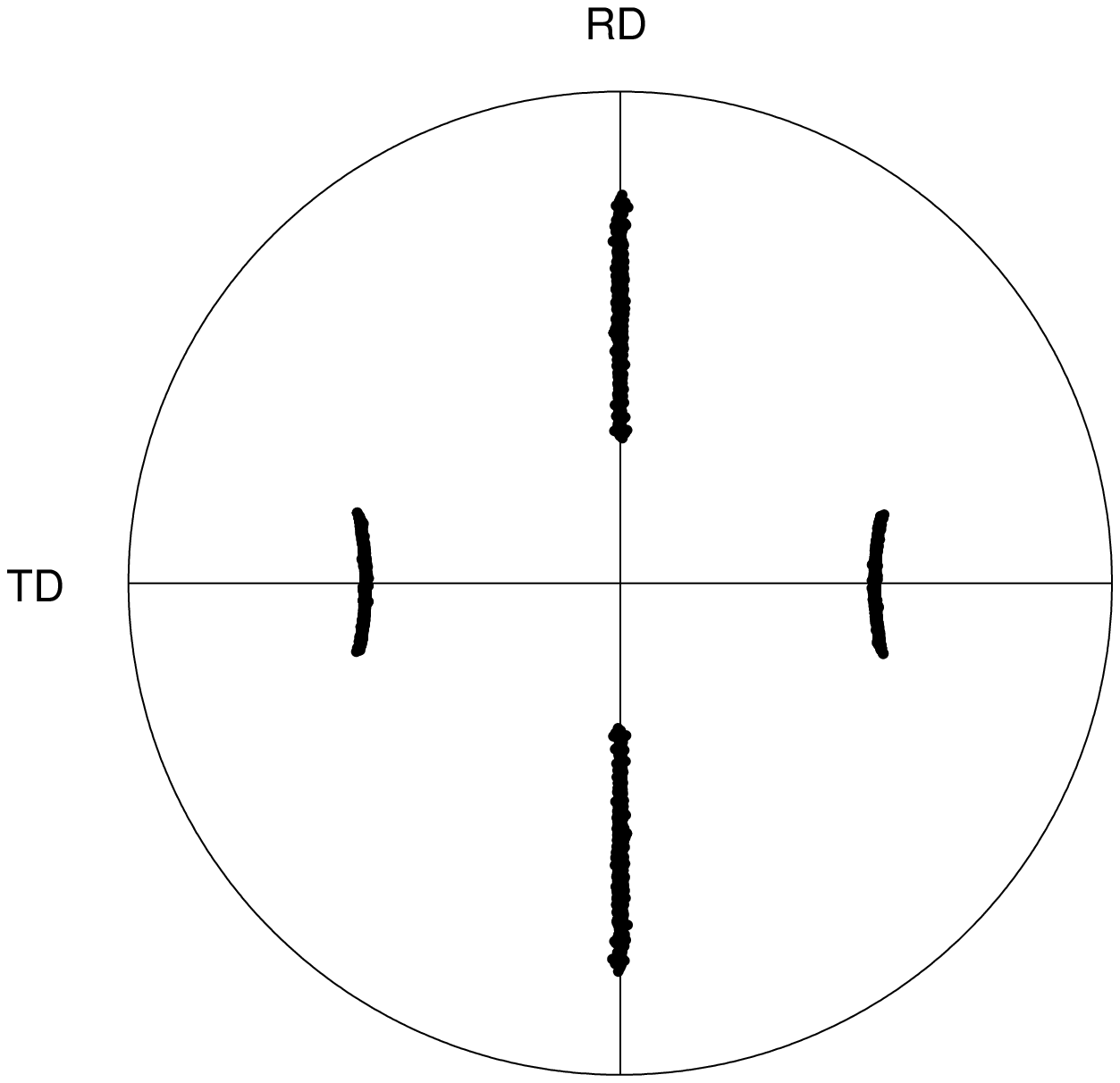}%
}
\subfigure[0.10]
{
\includegraphics[width=.45\textwidth, height=!] {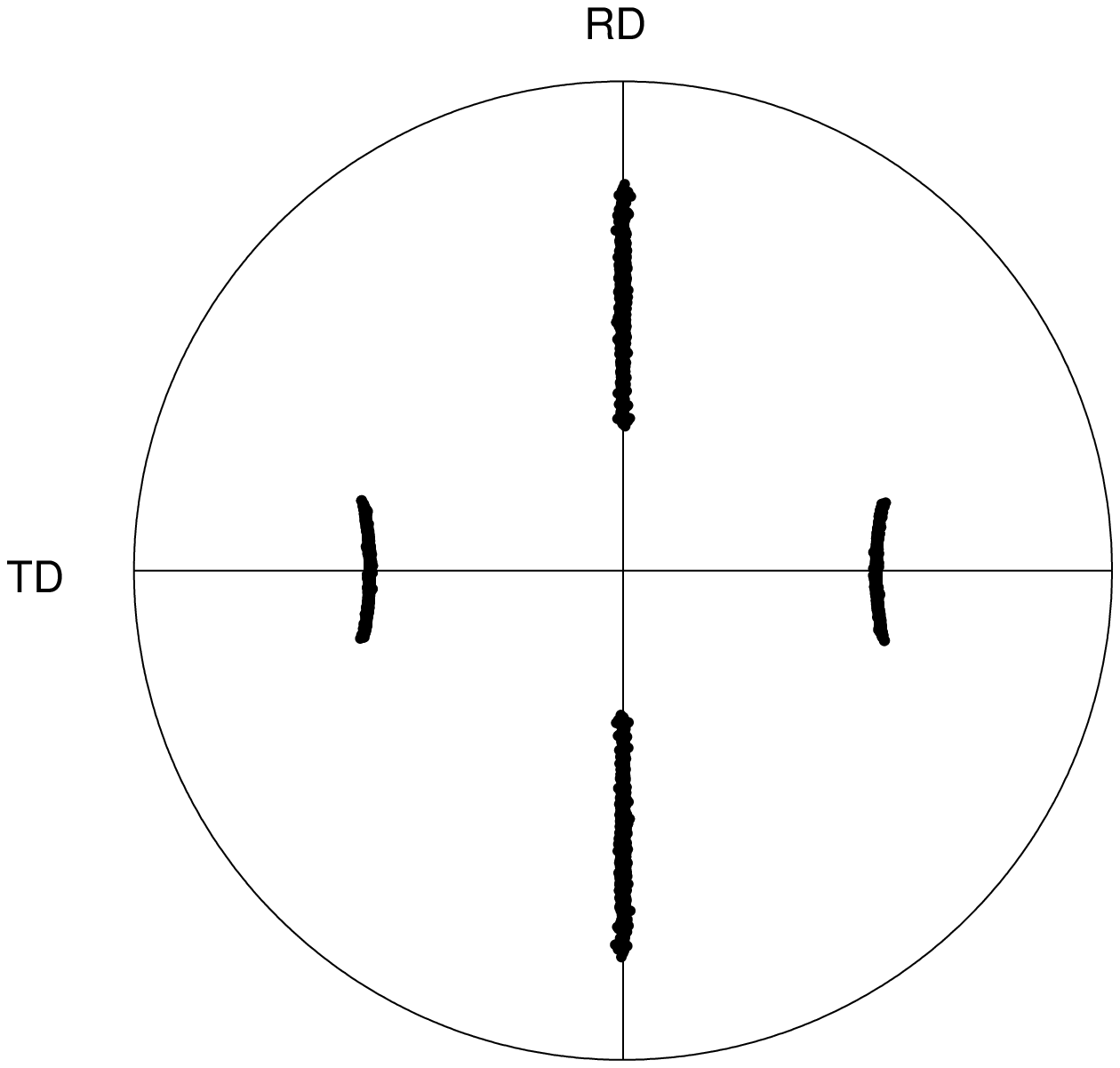}%
} \caption{Textures predicted by the implicit model simulations for
different time factors.} \label{final implicit 111}
\end{figure}

Another way in which we can check our subcycling implementation is by
examining the predicted material texture at the end of the deformation. Each
of our rolling tests was given the initial texture in Figure \ref{initial
texture}. Figure \ref{final explicit textures} shows that we recover similar
$<$%
111%
$>$%
textures for all the explicit simulations, given the same initial textures.
Also, the implicit model predicts similar textures to the explicit for the
same time step; see Figure .These textures agree well with measured textures
from the literature. In order to produce these textures, we reduced the
interaction parameter $a_{0}$ to $2\times10^{-5}$ (see Table \ref{fcccoef}).
Figure \ref{final implicit 111} shows the final textures obtained from
implicit model simulations, using time steps at 50\% and 10\% of the mesh
value. These textures are similar to those predicted by the explicit model.

\begin{figure}
[ptb]
\begin{center}
\includegraphics[width=.60\textwidth,height=!]{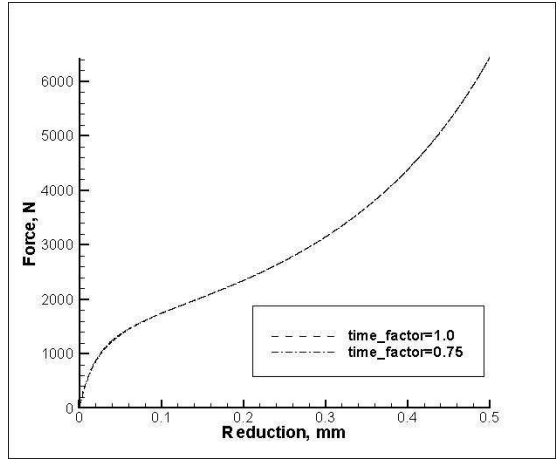}%
\caption{Force-deformation curves for two simulations using viscosity exponent
$m=1.0.$}%
\label{high visc force}%
\end{center}
\end{figure}

\begin{figure}
[ptb]
\begin{center}
\includegraphics[width=.60\textwidth,height=!]{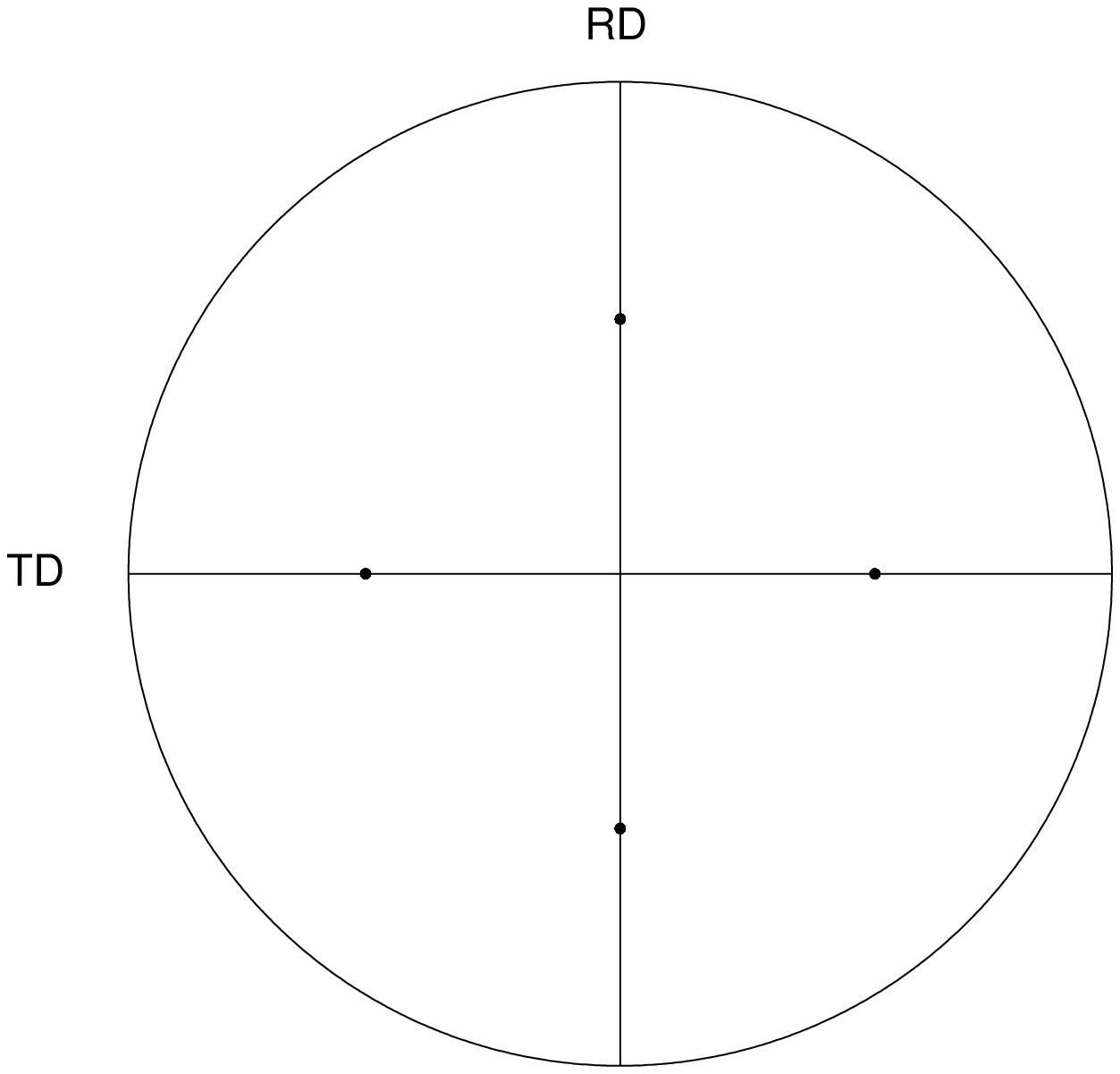}%
\caption{Texture prediction for the high-viscosity simulation.}%
\label{high visc texture}%
\end{center}
\end{figure}

Note that the predicted forces in the foregoing figures are quite small, on
the order of $10$ N. Since this force is applied to a $1$ mm cube, this
translates directly to $10$ MPa. The low force values returned are due to our
rather small choice of viscosity exponent $\left(  m=0.1\right)  $. This
exponent gives very good force matching for the quasi-static cases above, but
it does not reflect the physics of the dynamic simulations very well. This
viscosity exponent was chosen for these simulations to test the limitations of
our explicit model. High-viscosity deformations will have lower rates, thus
less plasticity and hardening, and therefore will easily converge for larger
times (i.e., $\Delta t_{c}$ will increase with the viscosity exponent). By
taking on the more challenging rate-independent limit, we can be sure our
model will perform well in the somewhat easier high-viscosity simulations. To
demonstrate this, we present Figure \ref{high visc force}, which shows results
for time\_factors of 1.0 and 0.75 with the viscosity exponent changed to 1.0.
The forces increase significantly from what we saw in the rate-independent
limit; now, our forces imply stresses in the range of several GPa.
Interestingly, the simulation using the full time step was 25\% (6 minutes)
faster than the simulation with 75\% of the mesh step for this set of
parameters. Since the larger viscosity exponent leads to less plasticity and
therefore less hardening, it follows that the $\Delta t_{c}$ values for these
simulations are longer, reducing the number of subcycles needed to converge.
Figure \ref{high visc texture} shows the texture for the case with
time\_factor=1.0; our agreement with measured values degrades. In fact, we
detect no texture evolution whatsoever, since the amount of plastic
deformation has reduced considerably due to the increased viscosity exponent.
This question, unfortunately, cannot be resolved at the present time. The
published texture data is measured after the completed deformation, after a
period of time long enough to allow the deformed sample to relax. Presumably,
the texture evolution continues after the test is completed. A more accurate
comparison would be drawn between the current results and \textit{in situ}
texture evolution data. The appropriate set of model parameters may then be determined.%

\begin{figure}
[ptb]
\begin{center}
\includegraphics[width=.60\textwidth,height=!]{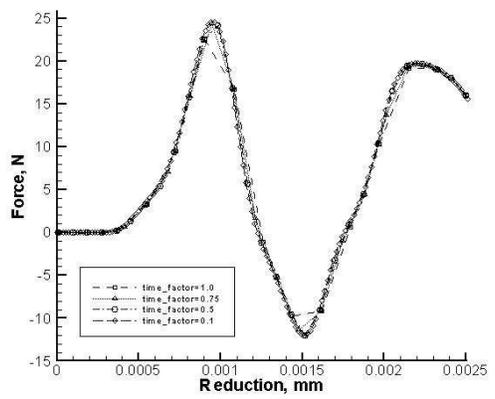}%
\caption{Close-up of the foregoing figure, with symbols designating points
returned.}%
\label{1x1x1force close-up}%
\end{center}
\end{figure}

\begin{figure}
[ptb]
\begin{center}
\includegraphics[width=.60\textwidth,height=!]{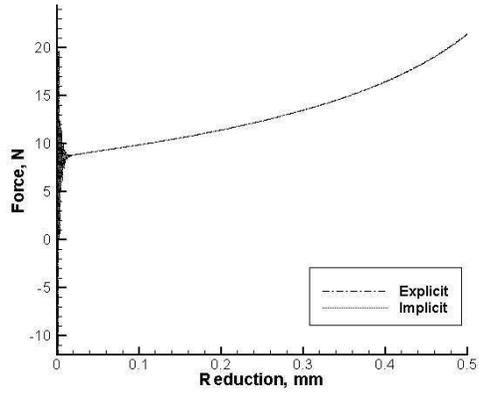}%
\caption{Force response predictions for the explicit and implicit model, both
using 50\% of the maximal mesh time step.}%
\label{implicit explicit}%
\end{center}
\end{figure}

\begin{figure}
[ptb]
\begin{center}
\includegraphics[width=.60\textwidth, height=!]{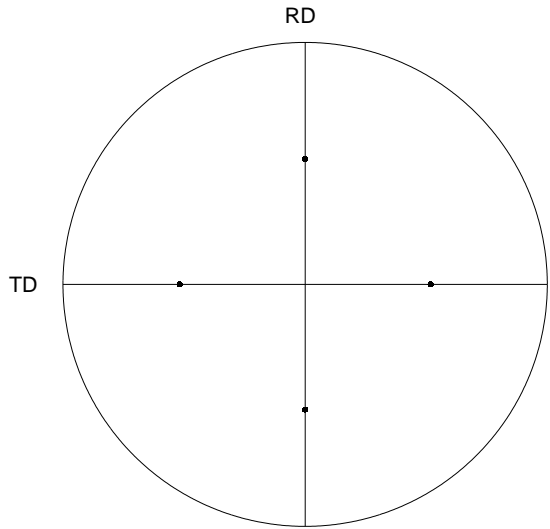}%
\caption{Initial $<$111$>$ texture for the tests in this section.}%
\label{initial texture}%
\end{center}
\end{figure}


\subsection{Polycrystal tests}

We have also performed some simulations of polycrystals under similar
conditions. We have used both Taylor averaging and direct numerical simulation
(DNS) to model the polycrystalline behavior based on the single-crystal model.
A discussion of the merits and drawbacks of these approaches has been
discussed elsewhere (see Zhao, \textit{et al }\cite{zhao04}). We provide some
of the results of these simulations here as they pertain to the explicit
constitutive update scheme, with and without subcycling.

We noted above the large fluctuations in force-deformation response
for a single integration point using the subcycling implementation.
These oscillations had little effect on the overall results for
single crystals; presumably their effects were damped out by the
overall bulk of the available quadrature points. However, we find
that these fluctuations are important for modeling of polycrystals.
Specifically, we tested a sample using the same mesh as the above,
but applied the deformation to a polycrystal having 91 orientations
using a Taylor averaging approach; that is, we presume the overall
deformation gradient and deformation rate are the same for all
orientations, and average to find the resulting stress. Where the
subcycling approach allowed our simulations to use the 100\% of the
time step allowable by the mesh for a single crystal (i.e.,
time\_factor=0.75), we were forced to reduce to a time step at 25\%
of that required by the mesh for this polycrystal. Simulations
applying DNS\ show similar behavior. We feel the difficulty is a
product of the interfaces of the polycrystalline grains. Instead of
having only a gradient between the part of the crystal that is
heavily deformed and the portion that is still nearly elastic (as in
the single-crystal case), the polycrystal introduces gradients
between parts of the crystal that have similar overall deformation
gradients and yet different plastic behavior. It stands to reason
that neighboring points in a polycrystalline mesh may well require
vastly different numbers of subcycles to meet the $\Delta t_{c}$
constraint. At points such as this, one element will then be on the
more stable converged path, while another is on the somewhat
volatile subcycling path. The result is that subcycling loses some
of its effectiveness for polycrystal modeling.

\begin{figure}
[ptb]
\begin{center}
\includegraphics[width=.45\textwidth,height=!]{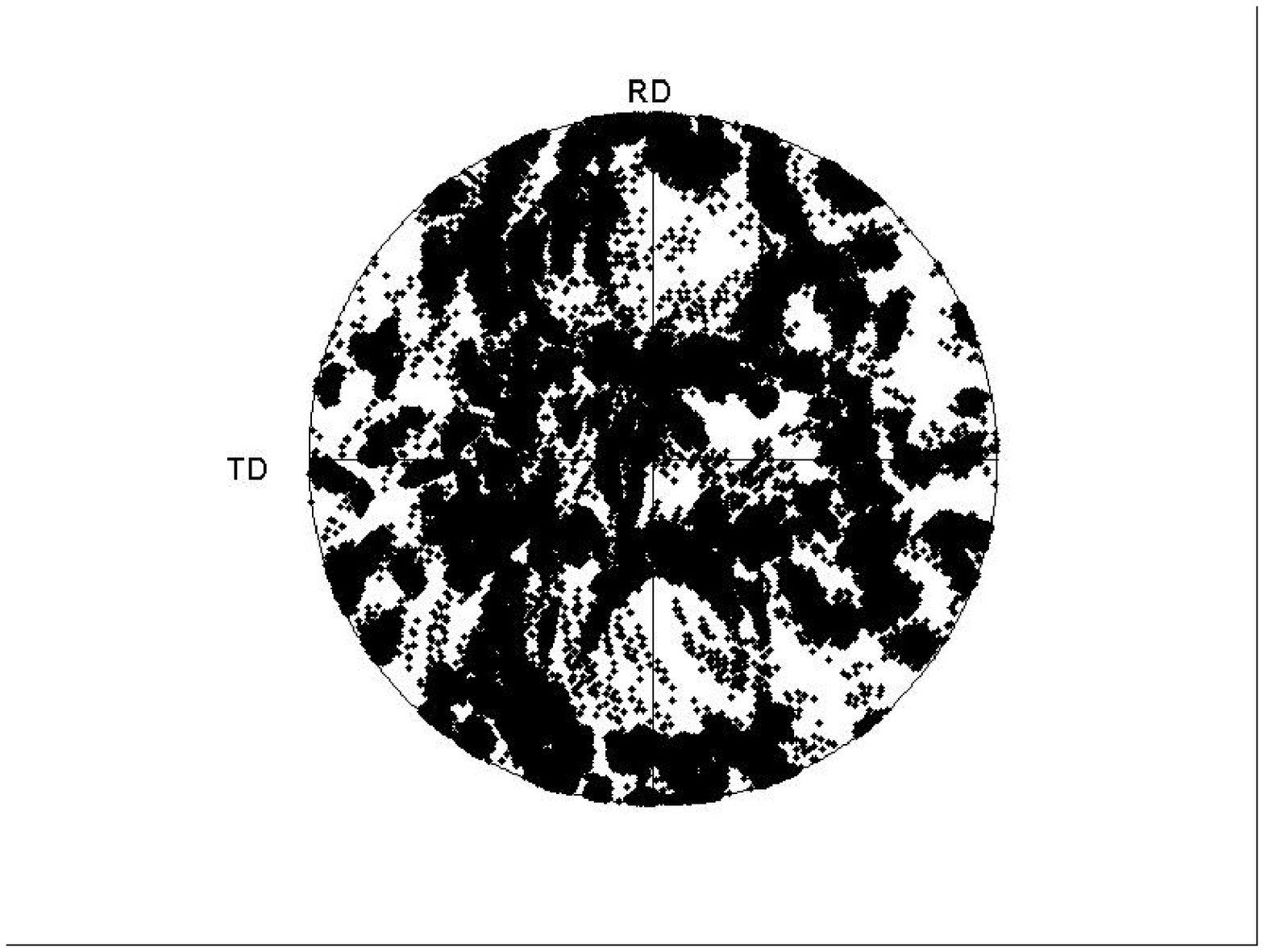}%
\caption{Predicted texture for Taylor averaging simulation using 91
orientations per integration point.}%
\label{taylor texture}%
\end{center}
\end{figure}

Subcycling still benefits the modeling effort, however. Instead of being
forced to a small step of 10\% the overall mesh step, we can still increase
our time step by a factor of 2.5. The results for averaging 91 orientations
took about four days, which is a linear scaling from the time taken by a
single orientation at the same time step. The results are shown below; the
force results in Figure \ref{taylor force}, and the final texture
in Figure \ref{taylor texture}.%

\begin{figure}
[ptb]
\begin{center}
\includegraphics[width=.45\textwidth,height=!]{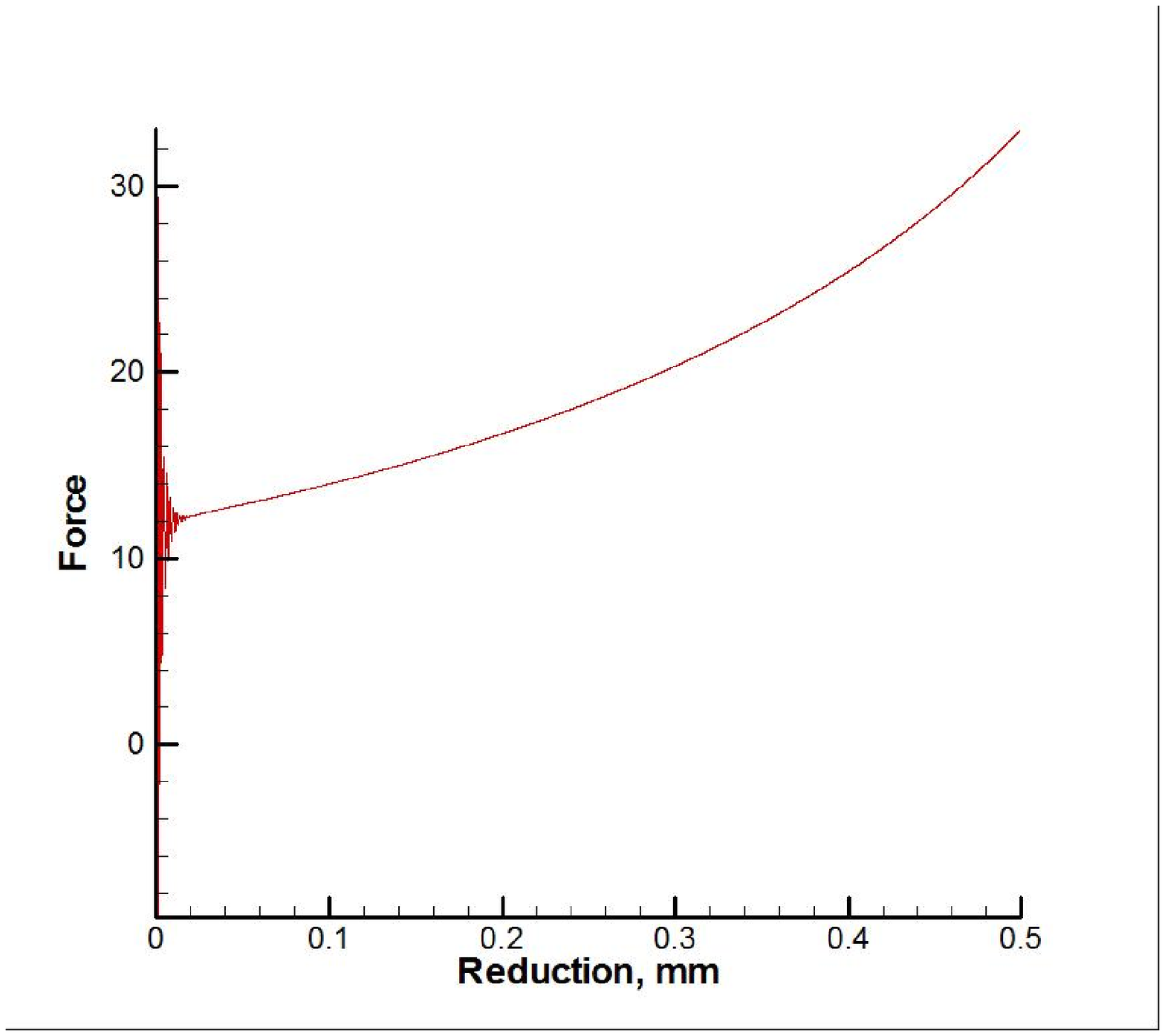}%
\caption{Predicted force-deformation curve for the polycrystal sample.}%
\label{taylor force}%
\end{center}
\end{figure}

\section{Conclusions}

We have demonstrated two model improvements which enhance the
computational speed of an implicit constitutive update, in cases
dominated by the global time step. Beginning with an implicit
formulation based on the model of Cuiti\~{n}o and Ortiz, we modified
the update to an explicit formulation. This explicit formulation was
shown to be five to six times faster than the original implicit
algorithm per update step at the constitutive level, without loss of
accuracy. The explicit update is limited by the size of time step
that can be taken. We were able to increase this maximal time step
by an order of magnitude for single crystals by introducing a
subcycling algorithm to the explicit form. All three models produce
the same stress-strain behavior at the integration point level for
the same input parameters.

We then applied our three update formulations to large-scale
finite-element calculations. The explicit update with subcycling was
shown to be able to integrate a larger time step for single crystals
than even the implicit, since the subcycling procedure changes the
larger input time step into several smaller steps that can be
evaluated by our explicit algorithm. For polycrystals, the
subcycling algorithm does not allow simulations to run at the full
mesh time step. However, the subcycling implementation allows an
increase in the maximum time step over the explicit model without
subcycling, and so results in an improvement in computational speed.

Notably, the implicit and explicit models each have their uses. The
implicit model is best suited to quasi-static simulations allowing
larger time steps to be used\footnote{Limited, of course, by the
converged time step}. In this sort of simulation, the gain of time
per step realized by the explicit model is countered by the number
of subcycles necessary to use the explicit model. Dynamic
simulations, for which the time step is typically very short, are
well-suited to the explicit model's capabilities. For such
simulations, the explicit model proves a robust algorithm.

\section*{Acknowledgements}

This work is sponsored by the U.S. Department of Energy's
Accelerated Strategic Computing Initiative (ASC) and the ASC Center
at the California Institute of Technology.

\bibliographystyle{abbrv}
\bibliography{algport}
\end{document}